\documentclass[twocolumn,showkeys,aps,prb,showpacs]{revtex4-1}
\usepackage{graphicx}
\usepackage{amsmath}
\usepackage[CJKbookmarks,dvipdfm,colorlinks,linkcolor=blue,citecolor=blue]{hyperref}

\begin{document}

\title{Nonmonotonic strain dependence of lattice thermal conductivity in  monolayer SiC: a first-principles study}

\author{San-Dong Guo and Jiang-Tao Liu}
\affiliation{School of Physics, China University of Mining and
Technology, Xuzhou 221116, Jiangsu, China}

\begin{abstract}
An increasing number of two-dimensional (2D) materials have already been achieved experimentally or predicted theoretically, which  have potential applications in nano- and opto-electronics. Various applications for electronic devices   are closely related to their
thermal transport properties. In this work, the strain dependence of phonon transport in monolayer SiC with a perfect planar
hexagonal honeycomb structure is investigated by solving  the linearized phonon Boltzmann equation. It is found that room-temperature lattice thermal conductivity ($\kappa_L$) of monolayer SiC is two orders of magnitude lower than that of graphene. The low $\kappa_L$ is due to small group  velocities and short  phonon lifetimes, which can also be explained by polarized covalent bond due to  large charge transfer from Si to C atoms. In considered strain range, it is proved that the SiC monolayer is mechanically and dynamically stable.
With increased tensile strain, the  $\kappa_L$ of SiC monolayer shows an unusual nonmonotonic up-and-down behavior, which is due to the competition between the change of phonon group velocities and phonon lifetimes of low frequency  phonon modes.  At
low strains ($<$8\%),  the  phonon lifetimes enhancement  induces the  increased $\kappa_L$, while at high  strains
($>$8\%) the reduction of group velocities  as well as the decrease of the phonon lifetimes are the major mechanism responsible for decreased $\kappa_L$.
Our works further enrich studies on phonon transports of 2D materials with a perfect planar
hexagonal honeycomb structure, and motivate farther experimental studies.

\end{abstract}

\pacs{72.15.Jf, 71.20.-b, 71.70.Ej, 79.10.-n ~~~~~~~~~~~~~~~~~~~~~~~~~~~~~~~~~~~Email:guosd@cumt.edu.cn}

\maketitle

\section{Introduction}
Due to their fascinating physical and chemical properties, 2D materials have attracted increasing attention since the successful
synthesis of graphene\cite{q1}. In comparison with the gapless graphene, semiconducting transition-metal dichalcogenide (TMD)\cite{q7}, group IV-VI\cite{q8}, group-VA\cite{q9,q10},  group-IV\cite{q11}, GaN\cite{q12} and ZnO\cite{q13}  monolayers   have intrinsic energy band gaps.  Thermal management is very important for next generation of electronics and
optoelectronic devices based on these 2D materials\cite{q14}, which has been hot spot  in
the field of materials. The thermal transports of  many 2D materials have been studied from a combination of  first-principles calculations and the linearized phonon Boltzmann equation\cite{q15,q16,q17,p5,p5-1,p5-2}.
The phonon transports of 2D orthorhombic group IV-VI compounds  (GeS,
GeSe, SnS and SnSe)  have been systematically investigated, and they show diverse anisotropic properties along the zigzag and armchair directions\cite{q15}.
Phonon transport properties of 2D group-IV materials have been performed, and although the $\kappa_L$  decreases monotonically from  graphene to silicene to germanene, unexpected higher $\kappa_L$  is observed in stanene\cite{q16}. The thermal transports of group-VA elements (As, Sb, Bi) monolayers with graphenelike buckled structure have been studied, including both electron and phonon parts\cite{q17}. The phonon transports of  TMD $\mathrm{MX_2}$ (M=Mo, W, Zr and Hf; X=S and Se) monolayers have been systematically investigated\cite{p5}. The $\kappa_L$
of 2H-type TMD monolayers  are generally higher than those of  1T-type
ones due  to the larger acoustic-optical frequency gap\cite{p5}. Strain effects on $\kappa_L$ have also been carried out in various kinds of 2D materials, such Sb and AsP monolayer\cite{q18,q18-1}, 2D group-IV\cite{q19}, 2D $\mathrm{MoTe_2}$\cite{q19-1}  and 2D penta-structures materials\cite{q20}.  The $\kappa_L$ shows  diverse strain dependence, such as  monotonously increasing,  up-and-down and  monotonously decreasing behaviors with increasing strain.
\begin{figure}
  \includegraphics[width=5.5cm]{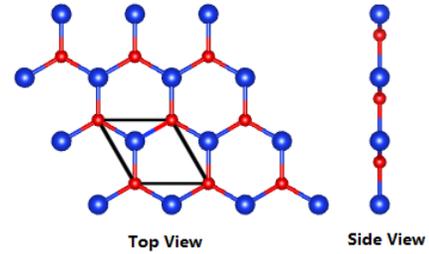}
  \caption{(Color online)The top and side views of monolayer SiC, and the frame surrounded by a black box is  unit cell. The blue and red balls represent Si and C atoms, respectively.}\label{st}
\end{figure}

Like graphene,  ZnO and GaN  monolayers possess a perfect planar
hexagonal honeycomb structural configuration, and their $\kappa_L$ have been investigated from a first-principles study\cite{q21,q22}.
The room-temperature $\kappa_L$  of monolayer ZnO is 4.5  $\mathrm{W m^{-1} K^{-1}}$ with the thickness of 3.04 $\mathrm{{\AA}}$, and it's lattice thermal conductivity shows anomalous temperature dependence\cite{q21}. The  $\kappa_L$  of monolayer GaN (300 K) is 14.93 $\mathrm{W m^{-1} K^{-1}}$ with the thickness of 3.74 $\mathrm{{\AA}}$, and the low  $\kappa_L$ can be explained by the special $sp$ orbital hybridization mediated by the Ga-$d$ orbital\cite{q22}. Recently, atomic resolution scanning transmission electron microscopy observations provide  direct experimental
indication of a two-dimensional form of silicon carbide, and the ground state of 2D SiC
is indeed completely planar by extensive simulations\cite{q23}. Similar to ZnO and GaN monolayers, 2D-SiC adopts a perfect planar
hexagonal honeycomb structure with the gap of 2.58 eV using  GGA. 

In this work, strain-dependent phonon transport properties of
SiC monolayer are studied  by solving the linearized phonon Boltzmann equation based on first-principles calculations.  The calculated room-temperature sheet thermal conductance of  SiC monolayer is  301.66 $\mathrm{W K^{-1}}$, which is substantially lower than that of
graphene (about 12884 $\mathrm{W K^{-1}}$)\cite{2dl}. The mode level phonon group velocities and  phonon lifetimes are used  to investigate
the mechanism underlying the lower  $\kappa_L$ of monolayer SiC compared with  graphene. The strongly
polarized Si-C bond, caused by large charge  transfer between Si and C atoms, induces large phonon
anharmonicity,  and gives rise to the intrinsic low $\kappa_L$ of monolayer SiC.
As the strain increases, the $\kappa_L$ of SiC monolayer  shows a  nonmonotonic up-and-down behavior, which can be understood by the competition between the change of phonon group velocities and phonon lifetimes of low frequency  phonon modes.

The rest of the paper is organized as follows. In the next
section, we shall give our computational details about phonon transport. In the third section, we shall present strain-dependent phonon transport of monolayer SiC. Finally, we shall give our conclusions in the fourth section.

\section{Computational detail}
First-principles calculations are carried out using the projected augmented wave (PAW) method, and the exchange-correlation  functional
of generalized gradient approximation of the Perdew-Burke-Ernzerhof (GGA-PBE) is adopted,
as implemented in the VASP code\cite{pv1,pv2,pbe,pv3}.
A plane-wave basis set is employed with
kinetic energy cutoff of 700 eV, and the $2s$ ($3s$) and  $2p$ ($3p$) orbitals of C(Si) atoms are treated as valance ones.
The unit cell  of monolayer SiC   is built with the vacuum region of larger than 17 $\mathrm{{\AA}}$ to avoid spurious interaction.
The electronic stopping criterion is $10^{-8}$ eV.
The  $\kappa_L$ of  monolayer SiC   is calculated by solving linearized phonon Boltzmann equation with the single mode relaxation time approximation (RTA),   as implemented in the Phono3py code\cite{pv4}.
The $\kappa_L$ can be expressed as:
\begin{equation}\label{eq0}
    \kappa=\frac{1}{NV_0}\sum_\lambda \kappa_\lambda=\frac{1}{NV_0}\sum_\lambda C_\lambda \nu_\lambda \otimes \nu_\lambda \tau_\lambda
\end{equation}
where $\lambda$, $N$ and  $V_0$ are  phonon mode, the total number of q points sampling Brillouin zone (BZ) and  the volume of a unit cell, and  $C_\lambda$,  $ \nu_\lambda$, $\tau_\lambda$   is the specific heat,  phonon velocity,  phonon lifetime.
The phonon lifetime $\tau_\lambda$ can be attained  by  phonon linewidth $2\Gamma_\lambda(\omega_\lambda)$ of the phonon mode
$\lambda$:
\begin{equation}\label{eq0}
    \tau_\lambda=\frac{1}{2\Gamma_\lambda(\omega_\lambda)}
\end{equation}
The $\Gamma_\lambda(\omega)$  takes the form analogous to the Fermi golden rule:
\begin{equation}
\begin{split}
   \Gamma_\lambda(\omega)=\frac{18\pi}{\hbar^2}\sum_{\lambda^{'}\lambda^{''}}|\Phi_{-\lambda\lambda^{'}\lambda^{''}}|^2
   [(f_\lambda^{'}+f_\lambda^{''}+1)\delta(\omega
    -\omega_\lambda^{'}-\\\omega_\lambda^{''})
   +(f_\lambda^{'}-f_\lambda^{''})[\delta(\omega
    +\omega_\lambda^{'}-\omega_\lambda^{''})-\delta(\omega
    -\omega_\lambda^{'}+\omega_\lambda^{''})]]
\end{split}
\end{equation}
in which $f_\lambda$  and $\Phi_{-\lambda\lambda^{'}\lambda^{''}}$ are the phonon equilibrium occupancy and the strength of interaction among the three phonons $\lambda$, $\lambda^{'}$, and $\lambda^{''}$ involved in the scattering.

The interatomic force constants (IFCs) are calculated by the finite displacement method.
 The second-order harmonic IFCs
are calculated using a 5 $\times$ 5 $\times$ 1  supercell  containing
50 atoms with k-point meshes of 3 $\times$ 3 $\times$ 1. Using the harmonic IFCs, phonon dispersion of monolayer SiC can be attained, as implemented in Phonopy package\cite{pv5}.  The phonon dispersion determines the allowed three-phonon scattering processes, and further the  group velocity  and specific heat can be attained. The third-order anharmonic IFCs are calculated using a 4 $\times$ 4 $\times$ 1
supercells containing 32 atoms with k-point meshes of 4 $\times$ 4 $\times$ 1.
 Based on third-order anharmonic IFCs, the three-phonon scattering rate can be calculated, and  further   the phonon lifetimes can be attained. To compute $\kappa_L$, the
reciprocal spaces of the primitive cells  are sampled using the 120 $\times$ 120 $\times$ 1 meshes.

For 2D material, the calculated  $\kappa_L$  depends on the length of unit cell  along z direction\cite{2dl}.  The $\kappa_L$ should be normalized by multiplying $Lz/d$, in which  $Lz$ is the length of unit cell along z direction  and $d$ is the thickness of 2D material, but the d  is not well defined.   In this work, the length of unit cell (18 $\mathrm{{\AA}}$) along z direction is used as the thickness of  monolayer SiC. To make a fair comparison between various 2D monolayers, the thermal sheet conductance can be used, defined as $\kappa_L$ $\times$ $d$.
\begin{figure}
  \includegraphics[width=8cm]{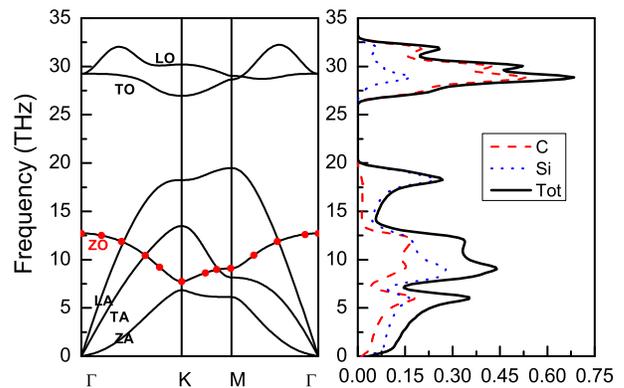}
  \caption{(Color online)Phonon band structures of monolayer SiC with the
corresponding density of states (DOS), and the atom partial DOS (PDOS) are also shown. The red dots represent ZO branch. }\label{ph}
\end{figure}

\begin{figure}
  \includegraphics[width=8cm]{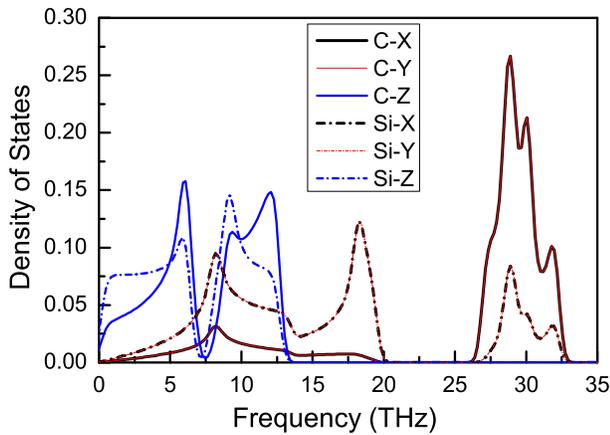}
  \caption{(Color online)The x, y and z components of atom PDOS of monolayer SiC. Due to hexagonal symmetry, the x and y components are the same.  }\label{ph1}
\end{figure}

\section{MAIN CALCULATED RESULTS AND ANALYSIS}
Like graphene,  monolayer ZnO and monolayer GaN, monolayer SiC possesses  a perfect planar
hexagonal honeycomb structure\cite{q23}, which can be constructed  by substituting one  C atom in
the unit cell of graphene with Si atom. The space symmetry group is $P\bar{6}M2$ for monolayer SiC, being
lower than that of graphene ($P6/MMM$), which is because monolayer SiC contains two kinds of atoms in the unit cell.
 The schematic crystal structure is shown in \autoref{st}, and the optimized lattice parameter is
3.104 $\mathrm{{\AA}}$.  It is worth noting that the bulk
SiC possesses buckled structure for SiC layer, being different from  planar structure for monolayer SiC, which may be  due to the variation of orbital hybridization from $sp^3$ to $sp^2$.

Based on the harmonic IFCs, the phonon dispersion of
 monolayer SiC is obtained along high-symmetry path, which along with total and partial density of states (DOS) are plotted in \autoref{ph}.
The phonon dispersion gives no imaginary frequencies, which indicates the
thermodynamic stability of monolayer SiC. Due to two atoms per unit cell, the phonon dispersion of monolayer SiC includes 3
acoustic and 3 optical phonon branches. The two highest
phonon branches are the in-plane transverse optical
(TO) and the in-plane longitudinal optical (LO) branches. It is clearly seen that there is a phonon band gap of 7.47 THz, separating TO and LO branches from
out-of-plane optical (ZO), in-plane longitudinal acoustic (LA), in-plane
transverse acoustic (TA) and out-of-plane acoustic (ZA) branches. The phonon band gap may be caused by different atomic masses of C and Si atoms.
Based on the highest acoustic frequency, the Debye
temperature can be attained by $\theta_D=h\nu_m/k_B$, where $h$ is the Planck constant, and $k_B$ is the Boltzmann
constant. The calculated value is about 935 K, which is lower than that of graphene (1977 K)\cite{q22}.
 The ZO branch crosses with the TA and LA branches, and there is a phonon band gap of 0.89 THz between ZA and ZO branches.
 Similar crosses can also be found in ZnO and GaN monolayers\cite{q21,q22}, but a phonon band gap is absent between ZA and ZO branches for monolayer GaN.
 The TA and LA branches are linear near the $\Gamma$ point, while the ZA branch
 deviates from linearity near the $\Gamma$ point, which can be found in many 2D materials\cite{q15,q16,q17,q21,q22}. The partial DOS
indicates that TO and LO  branches
are mainly contributed by the vibrations of
C atoms.  According to \autoref{ph1}, TA and LA  branches
are mainly from Si vibrations, while ZA and ZO branches are  contributed by the
vibrations of C and Si atoms.
\begin{figure}
  \includegraphics[width=8cm]{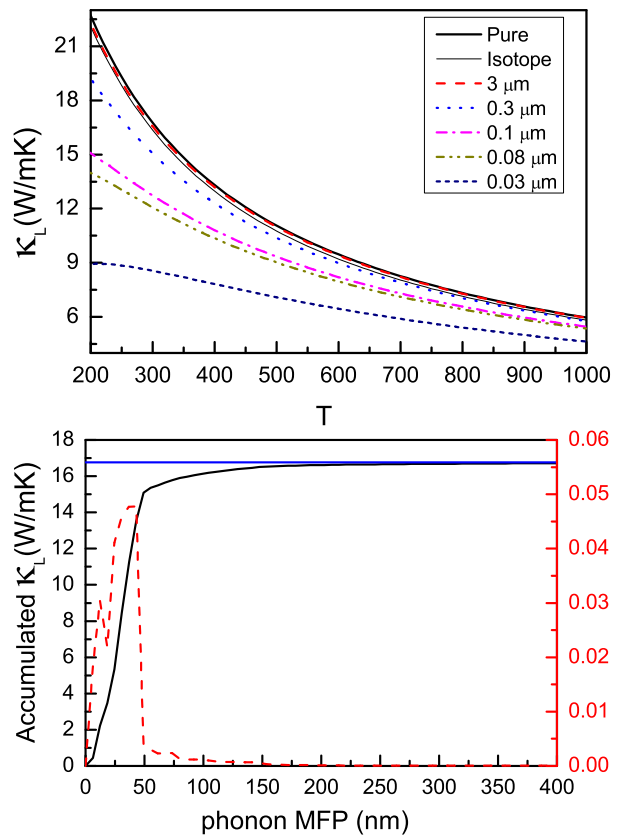}
  \caption{(Color online) Top: the lattice thermal conductivities  of infinite (Pure and Isotope) and finite-size (3, 0.3, 0.1, 0.08 and 0.03 $\mathrm{\mu m}$) monolayer SiC as a function of temperature;
   Bottom: cumulative lattice thermal conductivity of  infinite (Pure) monolayer SiC with respect to phonon mean free path at room temperature, and the derivatives.  The horizontal blue line represents room-temperature lattice thermal conductivity.}\label{kl}
\end{figure}
\begin{figure*}
  \includegraphics[width=16cm]{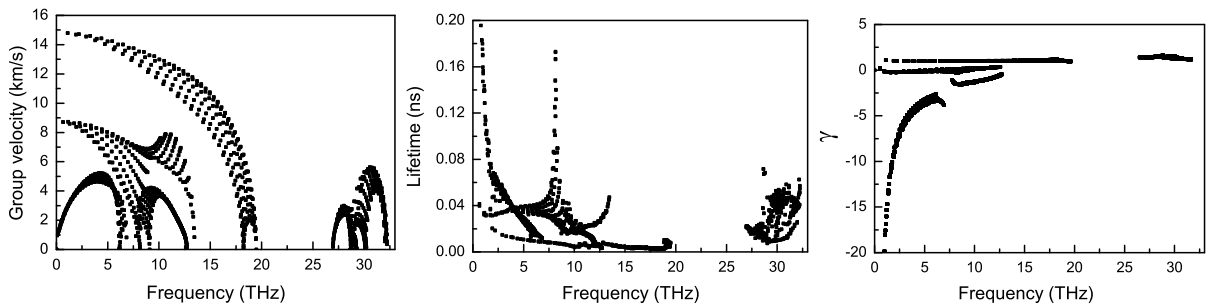}
  \caption{The mode level phonon group velocities, phonon lifetimes (300K) and   Gr$\mathrm{\ddot{u}}$neisen parameters  of infinite (Pure) monolayer SiC in the first BZ.}\label{v}
\end{figure*}
\begin{figure}
  \includegraphics[width=8.0cm]{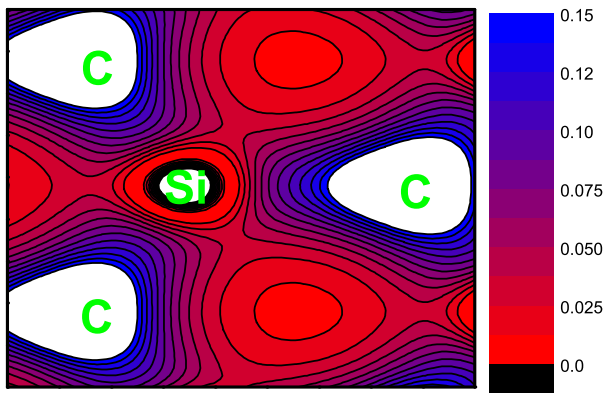}
  \caption{(Color online) The charge density distributions of monolayer SiC (unit:$\mathrm{|e|}$/$\mathrm{bohr^3}$).  }\label{den}
\end{figure}
\begin{table}[!htb]
\centering \caption{Born effective charges  $Z^*$ of Si and C atoms and the
dielectric constants ($\varepsilon$) of monolayer SiC.  They along other directions are zero except  $xx$, $yy$ and $zz$ directions.}\label{tab1}
  \begin{tabular*}{0.48\textwidth}{@{\extracolsep{\fill}}cccc}
  \hline\hline
Direction & $Z^*$(C) &$Z^*$(Si)& $\varepsilon$ \\\hline\hline
$xx$ &-3.674&3.674& 2.531\\
$yy$ &-3.674&3.674& 2.531\\
$zz$ &-0.271&0.271&1.180\\\hline\hline
\end{tabular*}
\end{table}

\begin{figure}
  \includegraphics[width=7.0cm]{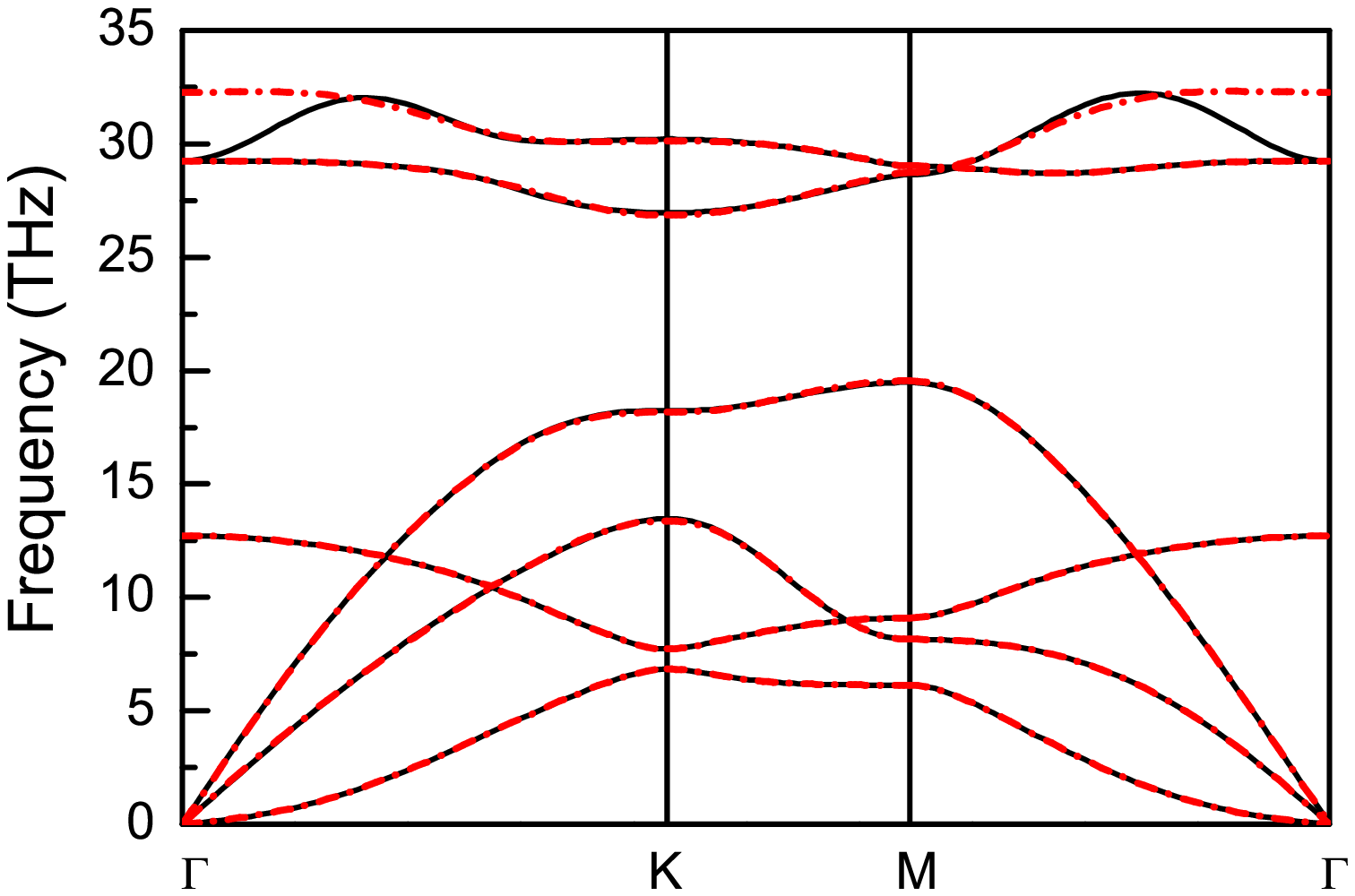}
  \includegraphics[width=7.2cm]{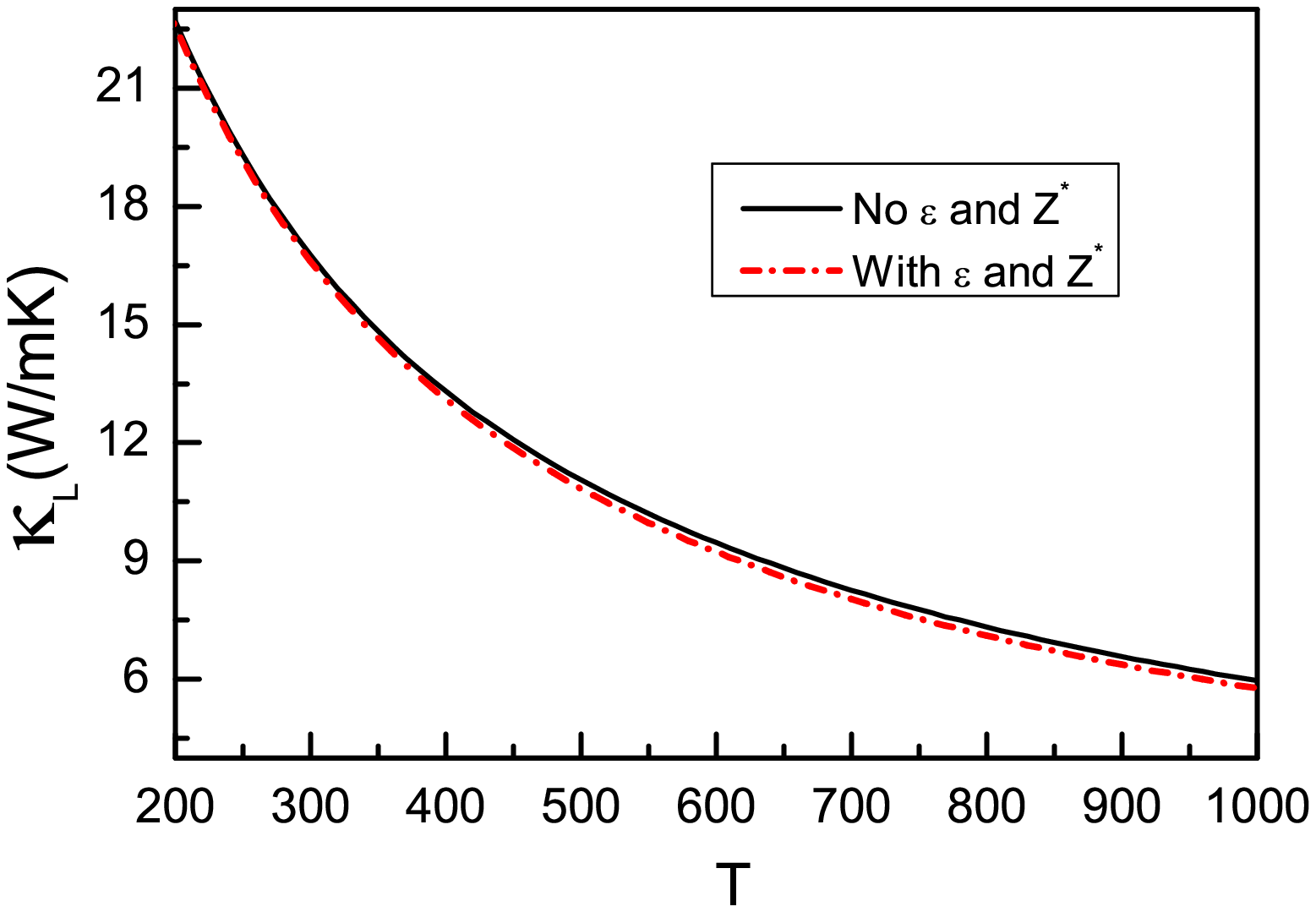}
  \caption{(Color online) The phonon band structures (Top) and lattice thermal conductivity (Bottom) of monolayer SiC with or without the dielectric constants $\varepsilon$ and Born effective charges $Z^*$.}\label{nac}
\end{figure}

The intrinsic $\kappa_L$ of monolayer SiC  is calculated by solving the linearized phonon Boltzmann equation within single-mode RTA method.
The phonon-isotope scattering  is calculated based on the formula proposed by Shin-ichiro Tamura\cite{q24}.
For boundary scattering, $v_g/L$ is just used as the scattering rate, where $v_g$ is the group velocity and $L$ is the boundary mean free path (MFP). The lattice thermal conductivities  of infinite (Pure and Isotope) and finite-size (3, 0.3, 0.1, 0.08 and 0.03 $\mathrm{\mu m}$) monolayer SiC as a function of temperature
are plotted in \autoref{kl}. The room-temperature $\kappa_L$ of infinite (Pure) monolayer SiC is 16.76  $\mathrm{W m^{-1} K^{-1}}$ with the thickness of 18 $\mathrm{{\AA}}$, and the corresponding thermal sheet conductance is 301.66 $\mathrm{W K^{-1}}$, which is  two orders of magnitude lower
than that for graphene (about 12884 $\mathrm{W K^{-1}}$)\cite{2dl}. It is clearly seen that the isotope
scattering has little effect on the lattice thermal conductivity of monolayer SiC, which may be due to  the strong phonon-phonon scattering.
With the sample size decreasing, the lattice thermal conductivity decreases due to enhanced boundary scattering. For the 3 and 0.3 $\mathrm{\mu m}$ cases,
the change is very small  with respect to infinite case.  The $\kappa_L$ for the  0.03 $\mathrm{\mu m}$ case is reduced to about half of that for infinite case at room temperature. To further understand the size dependence, the cumulative $\kappa_L$ along with the derivatives with respect to MFP (300 K) are  plotted in \autoref{kl}.
 With MFP increasing, the cumulative $\kappa_L$ approaches maximum. Phonons with MFP larger than 0.15 $\mathrm{\mu m}$ have very little contribution   to the $\kappa_L$. Phonons with MFP smaller than 0.03
$\mathrm{\mu m}$ contribute  around 50\% to the $\kappa_L$.

 \begin{figure}
  \includegraphics[width=8cm]{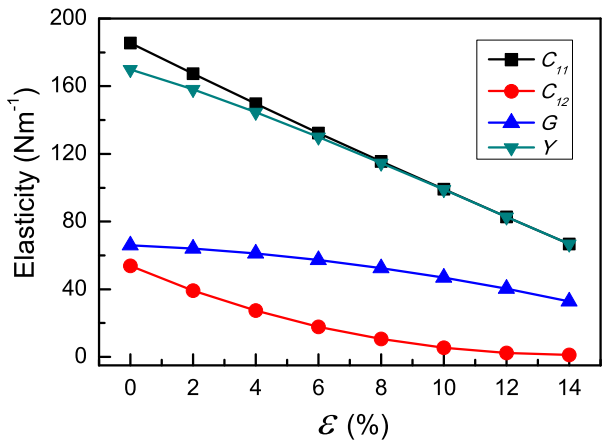}
  \caption{(Color online) The elastic constants $C_{ij}$, Young's moduli $Y$ and shear modulus $G$  vs strain for  SiC monolayer.}\label{elastic}
\end{figure}

\begin{figure}
  \includegraphics[width=8cm]{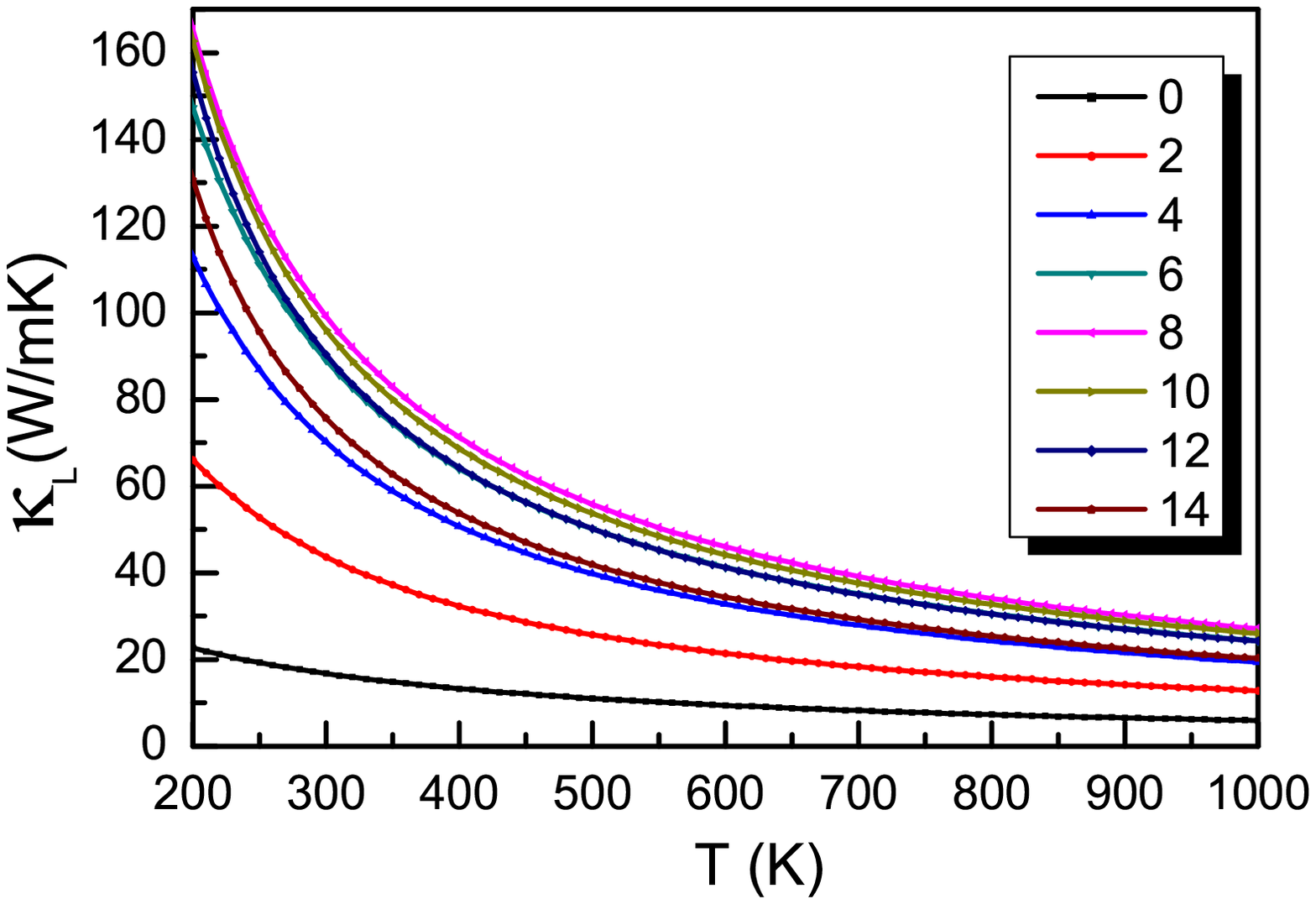}
   \includegraphics[width=8cm]{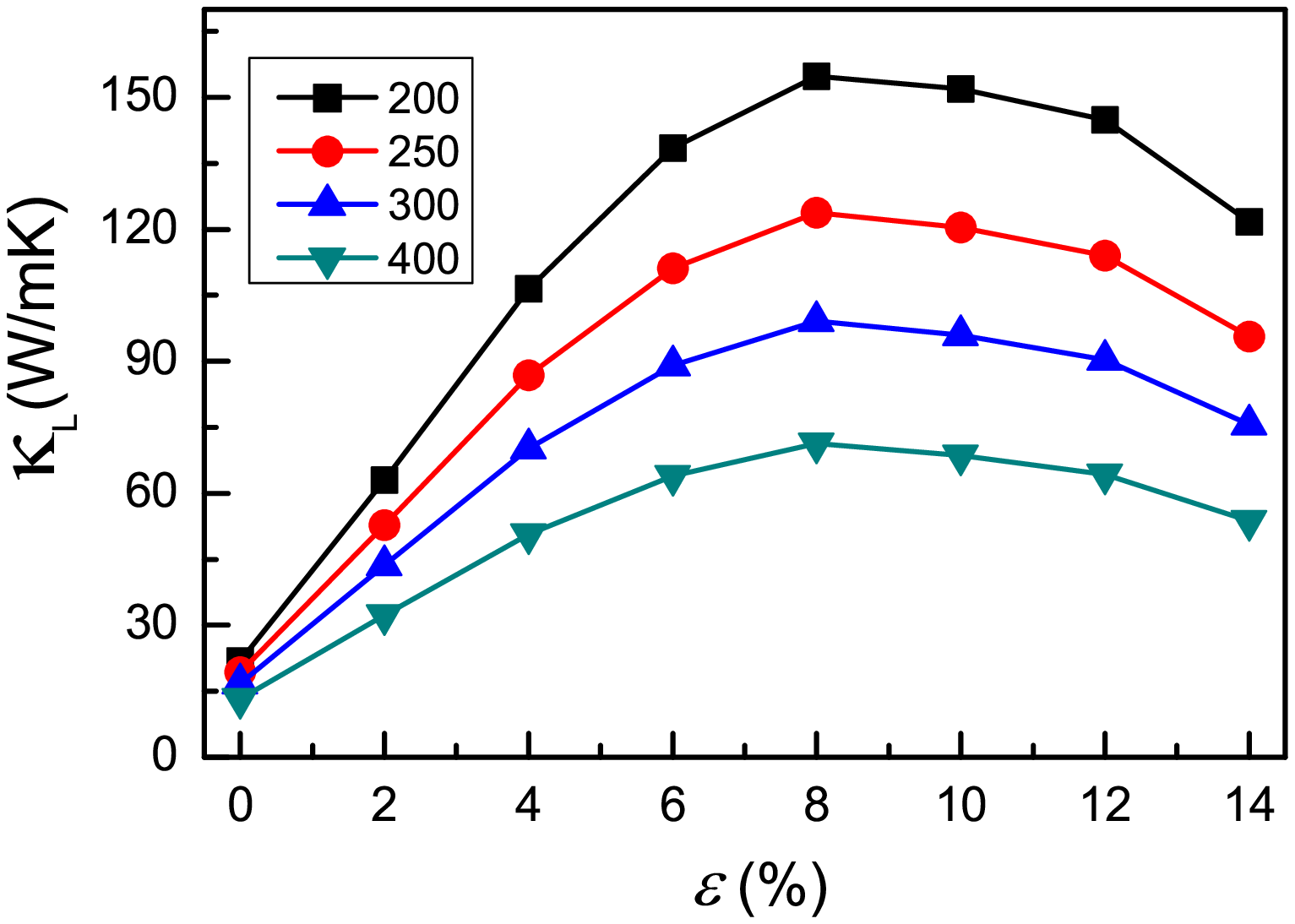}
  \caption{(Color online) Top: the lattice thermal conductivities  of SiC monolayer (0\%-14\% strain) as a function of temperature. Bottom: the lattice thermal conductivities (200, 250, 300 and 400 K)  of SiC monolayer  versus  strain.}\label{kl-c}
\end{figure}

\begin{figure*}
  \includegraphics[width=12cm]{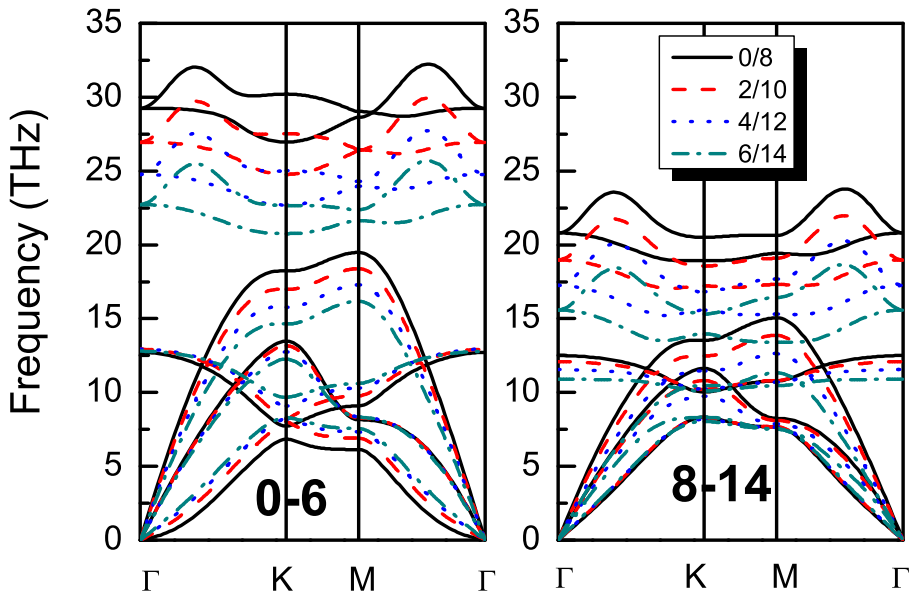}
  \caption{(Color online) The phonon dispersion curves of SiC monolayer with strain from 0\% to 14\%.}\label{ph-c}
\end{figure*}

\begin{figure}
  \includegraphics[width=8cm]{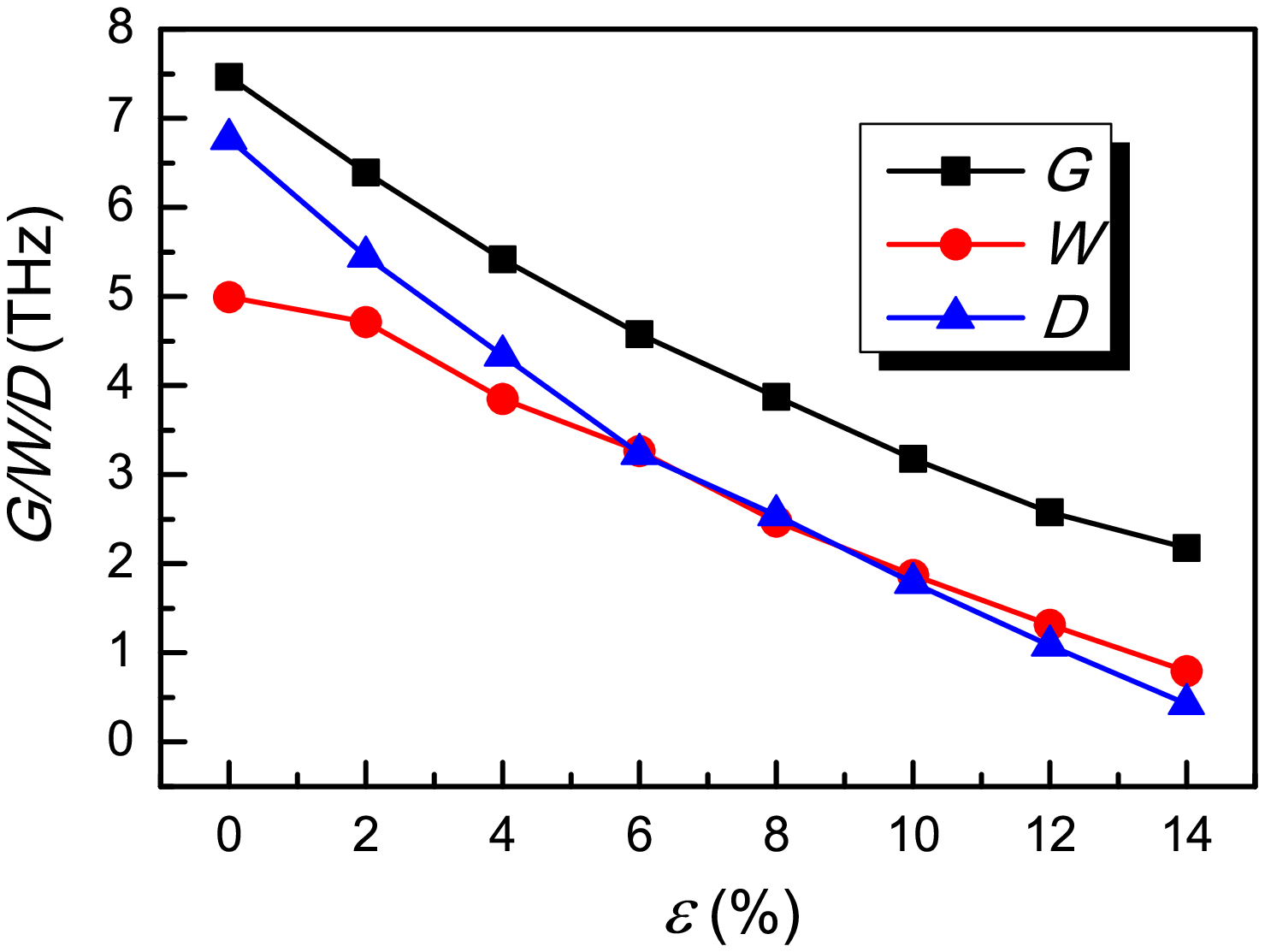}
  \caption{(Color online) The phonon gap $G$, the width of ZO mode $W$, and the difference of  maximum frequency between LA and ZO modes $D$  vs strain  for  SiC monolayer.}\label{g}
\end{figure}

To understand deeply phonon transport  of monolayer SiC,  the mode level phonon group velocities
and lifetimes are shown in \autoref{v}. The largest  group velocity  for  TA and LA branches near $\Gamma$ point is  8.74  $\mathrm{km s^{-1}}$ and 14.80  $\mathrm{km s^{-1}}$, which are lower than those of graphene\cite{q22}. Moreover, the overall phonon group velocities of monolayer SiC
are smaller than those of graphene, which partially leads to lower $\kappa_L$ for monolayer SiC than graphene.
The phonon lifetimes can be calculated from three-phonon scattering rate by third-order anharmonic IFCs.
It is found that ZA branch has relatively large phonon lifetimes. Due to the reflectional symmetry
of monolayer SiC, the  scattering
channels involving odd numbers of ZA modes are largely
suppressed, which leads to the low scattering rate of ZA, and then induces large phonon lifetimes. The lifetimes of  TO and LO phonon
branches  are much larger
than those below the gap, which can be explained by the weak phonon-phonon scattering caused by the large phonon gap.
The overall phonon lifetimes
of monolayer SiC are much smaller than those of  graphene\cite{q22}.  The strong coupling of ZO branch with acoustic branches is very important to
increase the scattering rate of the phonon modes below the gap by providing additional channels for the acoustic phonon
scattering, which can reduce phonon lifetimes of monolayer SiC.
The short phonon lifetimes together with small group velocities result in much lower
  $\kappa_L$ for  monolayer SiC than  graphene.

Mode Gr$\mathrm{\ddot{u}}$neisen parameters can be attained by third
order anharmonic IFCs, which  can reflect the strength of anharmonic interactions, determining the intrinsic phonon-phonon scattering.
The larger  $\gamma$ leads to lower  $\kappa_L$due to strong anharmonicity.
The  mode level phonon   Gr$\mathrm{\ddot{u}}$neisen parameters  of infinite (Pure) monolayer SiC in the first BZ are plotted in \autoref{v}.
For LA, TO and LO branches, the
$\gamma$ is fully positive.
For TA phonon modes, it shows both negative and partial positive $\gamma$.
The $\gamma$ is fully negative for ZA and ZO branches, where the large
negative  $\gamma$ of ZA branches shares the general feature of 2D materials
due to the membrane effect\cite{q25}. Although the ZA branch of monolayer SiC has larger $\gamma$ than that of some 2D materials,
the scattering of ZA is largely suppressed due to the symmetry-based selection rule.

The charge density can be used to describe the distribution of electrons
in real space, which is plotted \autoref{den} for monolayer SiC. It is clearly seen that the charge density increases from Si atom to C atom,  which
means that charge transfer is produced  between Si and C atoms
when the Si-C bond is formed. The charge transfer from Si to C atom induces the strongly polarized
covalent bond. For graphene, there is no charge transfer  due to  the same atom types to form bond. The strongly polarized covalent bond
can give rise to larger phonon anharmonicity, and induces stronger intrinsic phonon-phonon scattering, which leads to lower $\kappa_L$ of monolayer SiC with respect to graphene.

The dielectric constants $\varepsilon$ and Born effective
charges $Z^*$  effects on lattice thermal conductivity are considered, as given in \autoref{tab1}.  The strongly polarized covalent bond can also be characterized by large $Z^*$  and $\varepsilon$. The LO-TO splitting  at the Brillouin zone center is produced by long-range electrostatic Coulomb interactions, which can be clearly seen in \autoref{nac}. It is noted that the $\kappa_L$ only has a slight change
with or without $\varepsilon$ and $Z^*$. It is because LO and TO branches have little contribution to total $\kappa_L$.

Strain can effectively tune  $\kappa_L$  in many 2D materials\cite{q18,q18-1,q19,q19-1,q20}. The  biaxial strain can be described by defining
$\varepsilon=(a-a_0)/a_0$, in which $a_0$ is the unstrain lattice constant. Firstly,
the strain dependence of elastic constants $C_{ij}$, Young's moduli $Y$ and shear modulus $G$  in SiC monolayer  are plotted in \autoref{elastic}.   They all  monotonically decrease from 0\% to 14\% strain. In considered strain range, all the obtained elastic constants
are positive, confirming the mechanical stability in strained SiC monolayer. The $Y$  of the unstrained SiC monolayer
is  lower than those of h-BN and graphene\cite{ela}, but higher than one of $\mathrm{MoS_2}$\cite{ela-1}.
In the strain range of 0\% - 14\%, the $\kappa_L$ of  SiC monolayer  as a function of temperature are plotted in \autoref{kl-c}, together with  $\kappa_L$ vs strain at the temperature of  200, 250, 300 and 400  K. The $\kappa_L$ shows a nonmonotonic
trend with increased  strain, and first increases from 16.75 $\mathrm{W m^{-1} K^{-1}}$  (unstrained) to 99.15 $\mathrm{W m^{-1} K^{-1}}$  (at 8\% strain) then decreases to 75.61 $\mathrm{W m^{-1} K^{-1}}$  (at 14\% strain).
Similar up-and-down behavior is also found in penta-$\mathrm{SiC_2}$\cite{q20}, $\beta$-AsP\cite{q18-1} and bilayer graphene\cite{bg}.
The room-temperature $\kappa_L$  at  8\%  strain is  about 5.9 times that of the unstrained case.

To identify the underlying mechanism of strain-dependent up-and-down behavior of $\kappa_L$ in SiC monolayer,
the phonon dispersions of SiC monolayer with 0\%-14\% strain  are plotted \autoref{ph-c}. It is clearly seen that  there are  no imaginary frequencies in considered strain range,  indicating  that strained SiC monolayer is dynamically stable. With increasing strain,
the dispersions of both TA and LA modes are softened, producing
the reduced phonon group velocities.   The  dispersion  of ZA mode  near $\Gamma$ point is stiffened,  indicating the
improved  phonon group velocities. It is noted that the change of ZA mode at small strain is more obvious than one at large strain.
These  can also be observed  in penta-graphene\cite{q20}, AsP\cite{q18-1}  and $\mathrm{MoTe_2}$\cite{q19-1}.
With  strain increasing, the  quadratic nature of  ZA mode near the $\Gamma$ point disappears, which can also be found in AsP and $\mathrm{MoTe_2}$\cite{q18-1,q19-1}. Due to less strongly interacting between atoms with increased strain, the  dispersions of optical branches overall move toward low frequency.
The phonon gap $G$, the width of ZO mode $W$, and the difference of  maximum frequency between LA and ZO modes $D$ as a function of strain are plotted in \autoref{g}.
It is clearly seen that they all decrease with increased strain, which can produce important effects on phonon transports of SiC monolayer.
The large phonon gap hinders acoustic+acoustic$\rightarrow$optical (aao) scattering
 due to the energy conservation, which can enhance $\kappa_L$. The reduced $G$ with increased strain can give negative
contribution to $\kappa_L$.  The reduced $W$ and $D$ have positive  contribution to $\kappa_L$ due to reduced  scattering channels for acoustic phonon.

\begin{figure}
  \includegraphics[width=8cm]{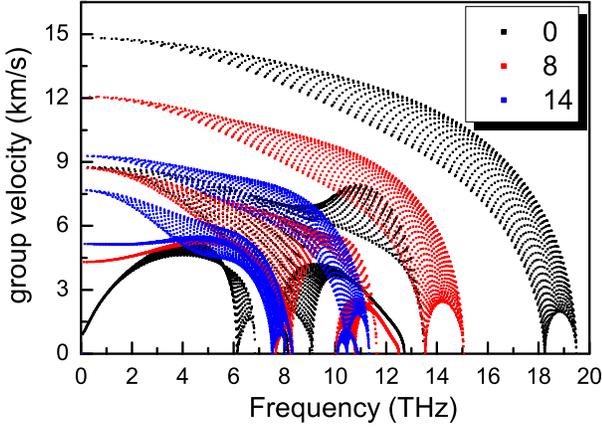}
  \caption{(Color online) The phonon mode group velocities of SiC monolayer  with strain   0\% (Black), 8\% (Red) and 14\% (Blue) in the first BZ for  modes below the phonon gap.}\label{v-c}
\end{figure}

\begin{figure}
  \includegraphics[width=8cm]{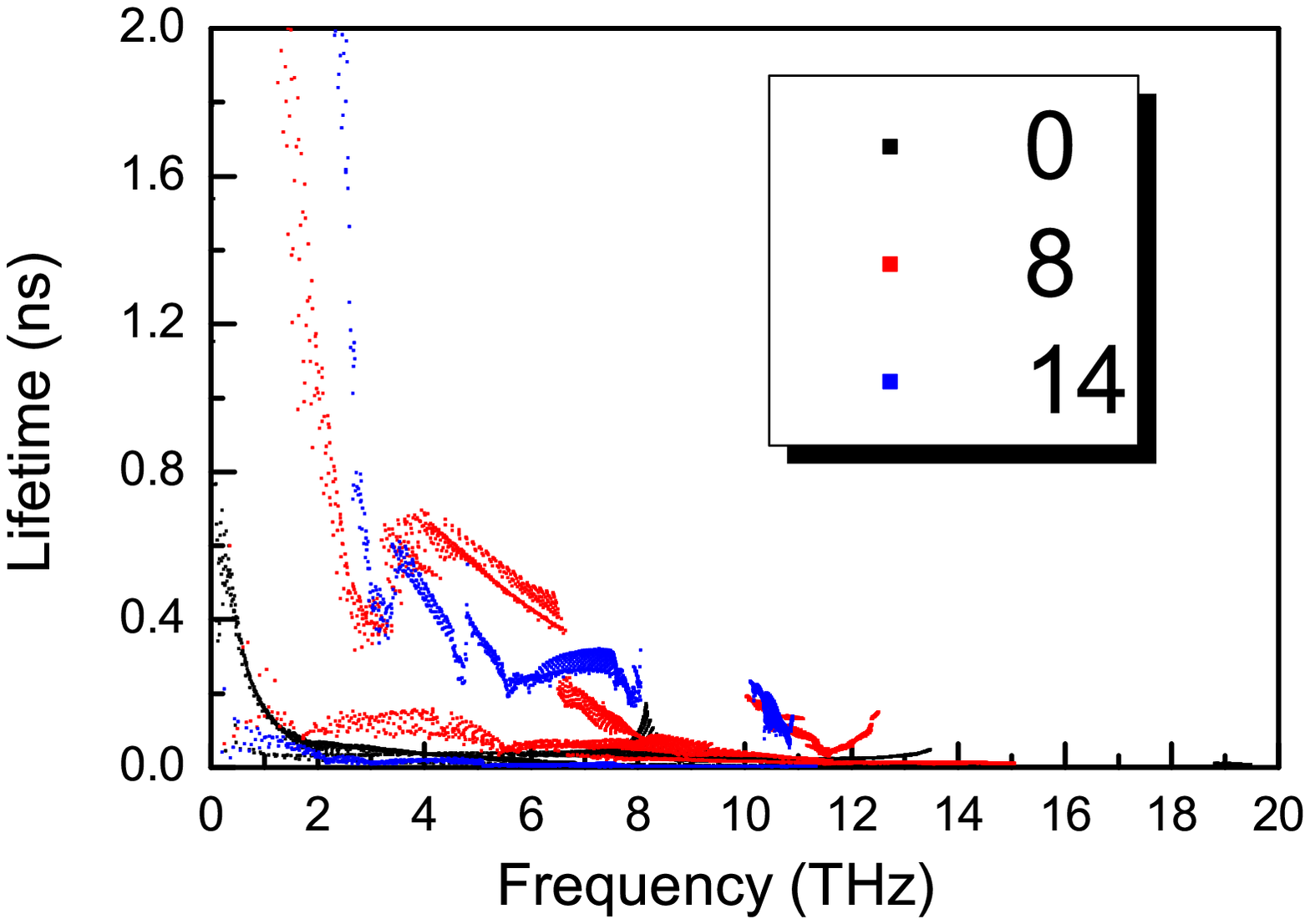}
    \includegraphics[width=8cm]{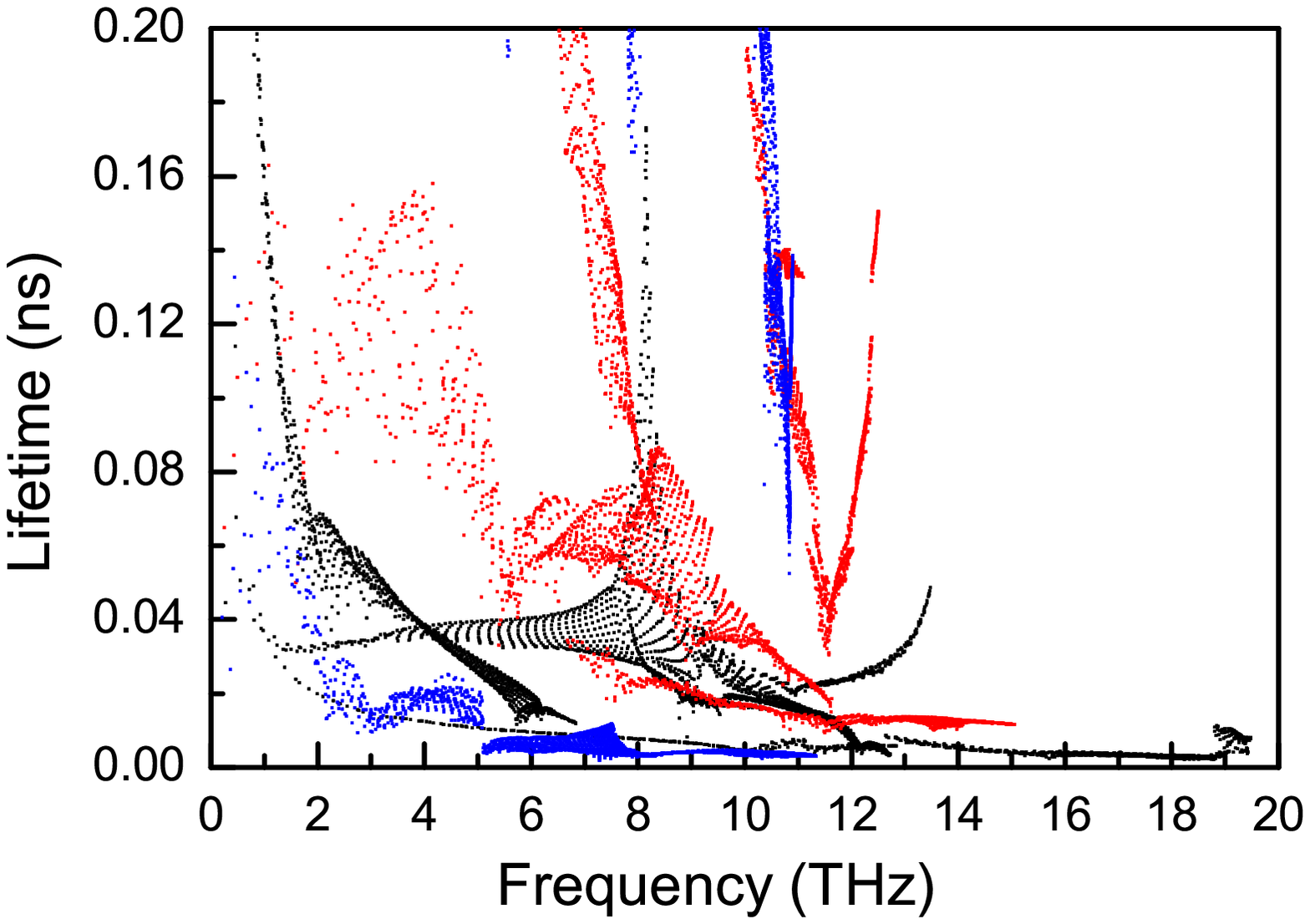}
  \caption{(Color online) The phonon mode lifetimes of SiC monolayer  with strain  0\% (Black), 8\% (Red) and 14\% (Blue) in the first BZ for  modes below the phonon gap.}\label{t-c}
\end{figure}

Due to dominant contribution to $\kappa_L$ from modes below the phonon gap, we only show these phonon mode group velocities and phonon lifetimes  with  0\%, 8\% and 14\% strains  in \autoref{v-c} and \autoref{t-c}. Due to stiffened ZA dispersion, group velocities  of ZA mode increase in low frequency region with increased strain. Near the $\Gamma$ point, the   group velocity  of ZA mode   increases from   0.94  $\mathrm{km s^{-1}}$ (0\%) to 4.30  $\mathrm{km s^{-1}}$ (6\%) to 5.15  $\mathrm{km s^{-1}}$ (10\%). In high frequency region, most of group velocities  of ZA mode increase from 0\% to 8\% strain, and have little change from 8\% to 14\% strain.  Due to softened phonon dispersions, the reduced  phonon group velocities of LA and TA branches are found with strain increasing. The phonon group velocities of ZO branch decrease with increased strain due to reduced $W$.  The reduction of most phonon group velocities  have negative
contribution to $\kappa_L$  with increased strain. It is clearly seen that most of phonon  lifetimes of SiC monolayer  firstly  increase, and then decrease.
The strain dependence of  phonon  lifetimes is consistent with that of $\kappa_L$.
The  opposite effects on  phonon lifetimes caused by  the reduce  of  between $G$ and $W$/$D$ lead to unusual strain dependence of phonon lifetimes.
Through considering synthetically strain dependence of phonon group velocities
and phonon lifetimes,   at
low strains ($<$8\%), the  phonon lifetimes  enhancement  is responsible for increased $\kappa_L$, while at high  strains
($>$8\%) the reduction of group velocities and phonon lifetimes   results in decreased $\kappa_L$.

\begin{figure}
  \includegraphics[width=8.0cm]{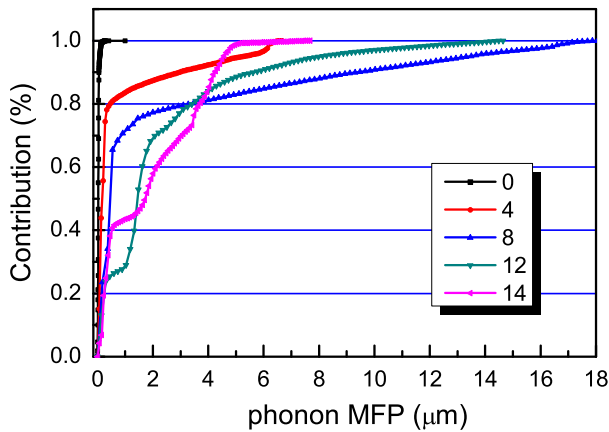}
  \caption{(Color online) At 0\%, 4\%, 8\%, 12\% and 14\% strains, the cumulative lattice thermal conductivity divided by total lattice thermal conductivity with respect to phonon MFP at room temperature.}\label{mfp-c}
\end{figure}

The  ratio between cumulative  and   total lattice thermal conductivity  of SiC monolayer with 0\%, 4\%, 8\%, 12\% and 14\% strains   as a function of phonon MFP are shown in \autoref{mfp-c}.  With increased  MFP, the ratio is close to one for all strains.
 The critical  MFP firstly increases with increased strain, and then decreases, which is identical with   $\kappa_L$.
 The critical  MFP changes  from 0.27 $\mu m$ (unstrained) to 17.87 $\mu m$ (at 8\% strain),  then  drop to 7.45 $\mu m$ (at 14\% strain).
The  critical MFP at 8\% strain  is 66 times larger than unstrain that, demonstrating  that strain can produce very  strong size effects for $\kappa_L$ of SiC monolayer. Similar results  can also be found in  antimonene, AsP monolayer, silicene, germanene and stanene\cite{q18,q18-1,q18-2}.

\section{Conclusion}
In summary, the first-principles calculations are performed  to predict the $\kappa_L$ of SiC monolayer under strain.
The calculated  room-temperature $\kappa_L$ of monolayer SiC is  substantially
lower than that of graphene. The underlying mechanism for the low   $\kappa_L$ of monolayer SiC
 can be understood by the mode level phonon group velocities and lifetimes. We
further perform analysis from the view of charge density distribution.
Significantly different from  that in graphene, there is very large charge transfer
between Si and C atoms, forming the strongly polarized covalent
Si-C bond. The strongly polarized Si-C bond gives rise to  the low  $\kappa_L$ of monolayer SiC by inducing large
phonon anharmonicity.
Within 14\% tensile strain, the
$\kappa_L$  of SiC monolayer first increases, and then decreases. The
maximum $\kappa_L$ is at  8\% tensile strain, which is  about 5.9 times that of the unstrained case.
This trend of $\kappa_L$ with increased strain  is mainly due to the strain-dependent phonon
lifetimes. Similar strain-dependent $\kappa_L$ may also be found in GaN and ZnO monolayers with the same perfect planar
hexagonal honeycomb structure.
Our works will motivate farther experimental studies, and studies of phonon
transports of other 2D materials with a perfect planar
hexagonal honeycomb structure.

\begin{acknowledgments}
This work is supported by the National Natural Science Foundation of China (Grant No.11404391).  We are grateful to the Advanced Analysis and Computation Center of CUMT for the award of CPU hours to accomplish this work.
\end{acknowledgments}

\end{document}